\documentclass[10pt,twocolumn,aps,prd,preprintnumbers,showpacs,superscriptaddress,nofootinbib,amsmath,amssymb,floats,floatfix,showkeys,notitlepage,longbibliography]{revtex4-1}

\usepackage{orcidlink}
\usepackage{comment}
\usepackage{lipsum}
\usepackage{graphicx}
\usepackage{subfigure}
\usepackage{palatino}
\usepackage{sans}
\usepackage{hyperref}
\hypersetup{colorlinks=true,linkcolor=blue,urlcolor=blue,citecolor=blue}
\usepackage[toc,page]{appendix}
\usepackage[normalem]{ulem}
\usepackage{adjustbox}
\usepackage{latexsym}
\usepackage{amsmath}
\usepackage{amssymb}
\usepackage{amsfonts}
\usepackage{dcolumn}
\usepackage{bm}
\usepackage{tikz}
\usepackage{bigints}
\usepackage{array,tabularx,multirow,booktabs}
\usepackage[tracking=true]{microtype}
\usepackage{soul} 
\usepackage{booktabs}
\usepackage{multirow}
\SetTracking{}{500}
\SetTracking{encoding={*}, shape=sc}{40}
\UseRawInputEncoding 
\allowdisplaybreaks

\begin{document} \sloppy
\title{Multimodal signatures of asymptotic (A)dS Kalb-Ramond black holes: Constraints through the shadow, weak deflection angle, and topological photon spheres}

\author{Reggie C. Pantig \orcidlink{0000-0002-3101-8591}} 
\email{rcpantig@mapua.edu.ph}
\affiliation{Physics Department, School of Foundational Studies and Education, Map\'ua University, 658 Muralla St., Intramuros, Manila 1002, Philippines.}

\author{Ali \"Ovg\"un \orcidlink{0000-0002-9889-342X}}
\email{ali.ovgun@emu.edu.tr}
\affiliation{Physics Department, Eastern Mediterranean University, Famagusta, 99628 North
Cyprus via Mersin 10, Turkiye.}

\begin{abstract}
This study explores novel static, neutral black hole solutions within Kalb-Ramond (KR) gravity in asymptotically (anti-)de Sitter [(A)dS] spacetimes, incorporating spontaneous Lorentz symmetry breaking (LSB) via an antisymmetric tensor field. Focusing on two metric configurations, we derive general analytical expressions for the horizon radius, photon sphere, shadow radius, and weak gravitational deflection angle. The universality of these expressions enables their applicability to a broad class of non-rotating spacetimes beyond KR gravity. By confronting these models with empirical data from the Event Horizon Telescope (EHT) and Solar System experiments, the work yields tight constraints on the Lorentz-violating parameter $\ell$, demonstrating that certain parameterizations - particularly Case B - yield observationally viable and physically consistent outcomes. Additionally, we employ topological methods to analyze black hole thermodynamics and photon sphere stability, uncovering critical geometrical structures in both thermodynamic and optical contexts. These multimodal signatures collectively offer a powerful framework for testing Lorentz-violating extensions of General Relativity (GR) and highlight the potential of topological diagnostics in gravitational physics.
\end{abstract}


\date{\today}
\maketitle

\section{Introduction} \label{intro}
The KR field is a rank-2 antisymmetric tensor field originally arising in string theory as a massless “B-field.” If such a field acquires a nonzero vacuum expectation value (VEV), it can spontaneously break local Lorentz symmetry, introducing preferred directions in spacetime. This idea of spontaneous LSB was first suggested in the late $1980$s in string theory contexts \cite{Yan_1983,Gasperini_1986,Kosteleck__1990}. In modern terms, the KR field with a VEV provides a concrete example within the Standard-Model Extension (SME) of gravity where an antisymmetric tensor background violates Lorentz invariance \cite{Altschul_2010}, laying the groundwork for modified gravity models that include a KR background. In such models, the KR field is typically coupled non-minimally to gravity (often through the Ricci curvature), ensuring that when the field condenses, it influences the spacetime geometry. This approach is analogous to \textit{bumblebee} gravity (where a vector field's VEV breaks Lorentz symmetry), but with a 2-form field instead 
\cite{Manton:2024hyc,Kostelecky:2003fs,Kostelecky:1988zi,Kostelecky:2000mm,Lessa:2020imi,Paul:2022mup,Lessa:2019bgi}. An important distinction is that the KR 2-form's field strength can be interpreted as spacetime torsion in some formulations, linking this framework to earlier torsion-based gravity theories. Notably, an early study by Kao et al. found that a KR field acting as \textit{hair} on a black hole tends to eliminate the event horizon, yielding a naked singularity for certain couplings \cite{Kao_1996}. Maluf and Neves derive exact static black-hole solutions in bumblebee gravity with a nonzero cosmological constant, demonstrating how spontaneous LSB modifies horizon structure \cite{Maluf:2020kgf}. Belchior and Maluf compute the one-loop radiative corrections in the bumblebee-Stueckelberg model, elucidating the quantum stability and renormalization of Lorentz-violating vector fields \cite{Belchior:2023cbl}. Belchior et al. extend KR bumblebee gravity by coupling to a global monopole, revealing novel spacetime geometries and topological charges in Ricci coupled Lorentz-violating frameworks \cite{Belchior:2025xam}. This suggested that naive insertions of a KR field could violate cosmic censorship, motivating more refined models. The contemporary KR gravity models improve on this by introducing the KR field's effects via controlled spontaneous LSB, resulting in well-behaved black hole solutions.

In KR gravity with Lorentz-violation, one typically adds to the Einstein–Hilbert action a term coupling the KR field $B_{\mu\nu}$ (or its 3-form field strength $H_{\lambda\mu\nu} = \partial_{[\lambda}B_{\mu\nu]}$) to curvature. A common choice is a contraction with the Ricci tensor, e.g. $\xi,B_{\mu}{}^{\alpha}B_{\nu\alpha}R^{\mu\nu}$, which effectively generates an extra stress-energy from the KR background. When $B_{\mu\nu}$ develops a constant background value (selecting a preferred frame), the Lorentz symmetry is locally broken, and the field equations admit deviations from GR parameterized by Lorentz-violation coupling constants. These deviations manifest as additional terms in the metric solution (often resembling “hair” analogous to electric charge or tidal charge). Importantly, the KR field is typically assumed non-dynamical or slowly varying in these solutions – it serves as a fixed background that backreacts on spacetime geometry. The result is a modified Schwarzschild or Kerr metric with extra parameters that measure the degree of LSB. The theoretical framework connects to broader modified gravity theories: for example, bumblebee gravity (with a vector field $B_\mu$) is a closely related Lorentz-violating model, and indeed many formalisms and phenomenological signals are analogous and studied \cite{Lessa:2019bgi,Nandi:2023vxt, Yang:2023wtu, Duan:2023gng, Masood:2024oej,Fathi:2025byw,Belchior:2025xam,Baruah:2025ifh,Duan:2023gng,Ortiqboev:2024mtk,Rahaman:2023swt,Yang:2023wtu,Sullivan:2020sop,Yan_1983,Alimova:2024bjd,Junior:2024vdk,Filho:2023ycx,al-Badawi:2024pdx,Bluhm:2004ep,Maluf:2015hda,Belchior:2023cbl,Lambiase:2023zeo,Li:2020dln,Guo:2023nkd,Delhom:2019wcm,Delhom:2020gfv,Filho:2022yrk,AraujoFilho:2024ykw,Altschul:2009ae,Assuncao:2019azw,Maluf:2018jwc,Liu:2024oas,Heidari:2024bvd}. The presence of Lorentz-violation effects is stringently restricted by high-precision observations within our Solar System, particularly when examined in the context of the bumblebee gravity model. Table \ref{tab:lorentz_constraints} summarizes constraints on the Lorentz-violating parameter, denoted by $\ell$, as recently determined by the authors in Ref. \cite{Yang:2023wtu}. These constraints were derived from detailed analyses of Solar System tests, including measurements of Mercury's orbital precession, gravitational light deflection, and the Shapiro time delay. Each of these tests places stringent limits on potential deviations from standard GR, thereby significantly narrowing the allowable parameter space for Lorentz-violating phenomena.
\begin{table}[]
    \centering
    \begin{tabular}{lc}
        \toprule
        Solar System tests & Bounds for $\ell$  \\
        \midrule
        Mercury precession & $-3.7\times 10^{-12}\leq \ell \leq 1.9\times 10^{-11}$ \\
        Light deflection & $-1.1\times 10^{-10}\leq \ell \leq 5.4\times 10^{-10}$ \\
        Shapiro time delay & $-6.1\times 10^{-13}\leq \ell \leq 2.8\times 10^{-14}$ \\
        \bottomrule
    \end{tabular}
    \caption{Lorentz-violating parameter $\ell$ from Solar System observations \cite{Yang:2023wtu}}
    \label{tab:lorentz_constraints}
\end{table}

Recent advancements in KR gravity have identified exact black hole solutions analogous to Schwarzschild and Schwarzschild-(A)dS metrics \cite{Liu:2024oas}. Nonetheless, a thorough exploration of more general static, neutral, and spherically symmetric black hole solutions remains relatively underdeveloped. Current advances in observational astrophysics, exemplified by gravitational wave detections by LIGO and Virgo and direct imaging of black hole shadows by the EHT \cite{EventHorizonTelescope:2019dse,EventHorizonTelescope:2019ths, EventHorizonTelescope:2022xqj,EventHorizonTelescope:2022wkp,EventHorizonTelescope:2022wok}, present unprecedented opportunities for testing gravitational theories beyond Einsteinian predictions. These observational breakthroughs strongly motivate further theoretical investigations into black hole solutions explicitly exhibiting Lorentz symmetry violations, potentially yielding clear observational signatures that can either validate or constrain alternative gravitational models.

Among the various observational techniques, the analysis of black hole shadows stands out as an exceptionally powerful method for directly probing spacetime geometry near black hole predictions \cite{Synge:1966okc,Luminet:1979nyg,Falcke:1999pj}. The shadow, defined by photon orbits at the photon sphere \cite{Claudel:2000yi,Virbhadra:2002ju,Virbhadra:1998dy,Virbhadra:2007kw,Virbhadra:1999nm,Virbhadra:2008ws,Virbhadra:2022iiy,Virbhadra:2002ju,Adler:2022qtb,Virbhadra:2022ybp,Zakharov:2014lqa,Zakharov:2023yjl}, contains critical information on possible deviations from GR predictions. Precision measurements, such as those obtained by the EHT \cite{EventHorizonTelescope:2019dse,EventHorizonTelescope:2019ths, EventHorizonTelescope:2022xqj,EventHorizonTelescope:2022wkp,EventHorizonTelescope:2022wok}, have provided stringent constraints and robust empirical tests for modified gravity theories \cite{Vagnozzi:2022moj,Bambi:2019tjh,Vagnozzi:2019apd,Khodadi:2020jij,Khodadi:2024ubi,Khodadi:2020gns,Battista:2023iyu,Wang:2025fmz,Uniyal:2023inx,Nozari:2023flq,Aliyan:2024xwl,Nozari:2024jiz,Campos:2021sff,Anacleto:2021qoe}. Complementing shadow observations, gravitational lensing, quantified by the weak deflection angle, offers sensitive probes of spacetime curvature \cite{Soares:2023err,Soares:2023uup,Soares:2024rhp,Soares:2025hpy,Javed:2023iih}, capable of detecting subtle Lorentz-violating effects otherwise undetectable by classical methods.

To be specific, several recent works in the context of bumblebee and KR gravity have deepened our understanding of how Lorentz-violating parameters influence black hole shadows, photon spheres, and deflection angles. Topological approaches have also matured in recent years. The topological classification of photon spheres has been used to distinguish black holes from naked singularities, particularly in (A)dS spacetimes. Wei (\cite{Wei:2020rbh}) demonstrates that each black hole spacetime admits a total topological charge of $−1$, corresponding to the presence of at least one unstable photon sphere, a result consistent with and reinforcing the present study’s topological analysis. In a comprehensive study of Schwarzschild-like black holes with a topological defect in Einstein–Hilbert–Bumblebee gravity, it was shown that both the global monopole and LSB parameters increase the radius of the black hole shadow and affect the deflection angle in the weak-field regime \cite{Gullu:2020qzu}. Moreover, the work by Araújo Filho et al. \cite{Filho:2023ycx} investigates how antisymmetric tensor effects influence black hole shadows and quasinormal modes, finding that LSB notably reduces both the photon sphere and the shadow radius. The inclusion of a cosmological constant was found to mitigate this reduction, suggesting an intricate interplay between dark energy and Lorentz-violating dynamics. In the context of KR gravity, recent studies have explored the shadow and weak gravitational lensing effects of Reissner–Nordstr\"om-like black holes, incorporating observational data from the EHT. These analyses revealed that an increase in the Lorentz-violating parameter leads to a decrease in shadow radius, offering potential constraints consistent with astrophysical observations \cite{Alimova:2024bjd}. Additionally, Hosseinifar et al. \cite{Hosseinifar:2024wwe} extended this topological perspective to charged AdS black holes within a KR background, showing how shadow radius, greybody factors, and deflection angles are modified by antisymmetric tensor fields and spontaneous LSB. Finally, the recent analytical estimation of the Lorentz-violating parameter $\ell$ using weak deflection angles and EHT data provides further confirmation of its observational viability. Lambiase et al. \cite{Lambiase:2024uzy} demonstrate that $\ell$ can be constrained from both Solar System and EHT data, emphasizing the robustness of black hole shadow analysis as a tool for probing Lorentz-violation. Other recent works in the context of bumblebee and KR gravity have deepened our understanding of how Lorentz-violating parameters influence black hole shadows, quasinormal modes, photon spheres, and deflection angles \cite{Kumar:2020hgm,Liu:2024lve,Pantig:2024ixc,Pantig:2024kqy,Mangut:2025gie,Hu:2025isj,AraujoFilho:2025fwd,Shi:2025rfq,Tang:2025eew,Ditta:2024lnb}.

This study is motivated by the need to probe LSB in the strong-gravity regime of black holes, where deviations from GR may yield observable signatures, particularly with recent advancements like the EHT imaging and high-precision Solar System tests. We investigate a novel (A)dS black hole solution within KR gravity, extending prior work \cite{Pantig:2024kqy} to derive universal analytical expressions for the shadow radius and weak gravitational deflection angle, applicable to any non-rotating black hole spacetime. By choosing KR gravity, we provide a concrete example of modified gravity that introduces subtle yet testable deviations from Einsteinian solutions through the antisymmetric tensor field’s VEV, enabling constraints on the Lorentz-violating parameter $\ell$ using EHT data and the Parameterized Post-Newtonian (PPN) formalism. Furthermore, our analysis of the thermodynamic topology of this solution illuminates phase transitions and the global structure of its parameter space, offering deeper insights into the theoretical and observational viability of gravitational extensions beyond the standard paradigm.

In Sect. \ref{sec2}, we briefly review the (A)dS black hole solution from KB gravity. In Sect. \ref{sec3}-\ref{sec5}, we derive the general analytic formula for the horizon, photon sphere, shadow radius, and weak deflection angle. The (A)dS black hole solution from KB gravity gives this opportunity to generalize the results from Ref. \cite{Pantig:2024kqy}. Finally, in Sect. \ref{sec6}, we investigate the topological thermodynamics and photon spheres of the (A)dS BH solution. We organized the paper as follows: Finally, we utilize $G = c = 1$ and the metric signature $(-,+,+,+)$.

\section{Brief review} \label{sec2}
In Ref. \cite{Liu:2024oas}, the authors analyze static, neutral, spherically symmetric black hole solutions in asymptotically (A)dS spacetimes within KR gravity, where spontaneous LSB arises from a nonminimally coupled antisymmetric tensor field $ B_{\mu\nu} $. The general static, spherically symmetric metric ansatz is given by
\begin{equation}
    ds^2 = -A(r)dt^2 + B(r)dr^2 + r^2 d\theta^2 + r^2 \sin^2 \theta d\phi^2,
\end{equation}
where $ A(r) $ and $ B(r) $ are metric functions determining the temporal and radial components of the spacetime geometry, respectively, to be solved via the field equations. These equations derive from the Einstein-Hilbert action augmented by a KR term coupling $ B_{\mu\nu} $ to the Ricci tensor and include a cosmological constant $ \Lambda $, which dictates the asymptotic (A)dS behavior ($ \Lambda > 0 $ for dS, $ \Lambda < 0 $ for AdS). A Lagrange multiplier field $ \lambda $ enforces the vacuum condition for $ B_{\mu\nu} $, ensuring its VEV, with norm $ b^2 $, is constant. The field equations are simplified using the relation $ B(r) = C_1 A(r)^{-1} $, where $ C_1 $ is a constant that distinguishes two solution classes.

For Case A ($ C_1 = 1 - \ell $), the metric function is
\begin{equation}
    A(r) = 1 - \frac{2M}{r} - \frac{(1 - \ell)(b^2 \lambda + \Lambda)}{3(1 + \ell)} r^2,
\end{equation}
where $ M $ is the black hole mass, $ \ell $ is the Lorentz-violating parameter measuring the strength of LSB, $ b^2 $ quantifies the KR field’s VEV magnitude, and $ \lambda $ is the Lagrange multiplier. The cosmological constant $ \Lambda $ relates to an effective cosmological constant $ \Lambda_e = \lambda / \xi_2 $, where $ \xi_2 $ is a coupling constant, via
\begin{equation} \label{e3}
    \Lambda = (1 - \ell) \Lambda_e,
\end{equation}
linking the asymptotic structure to Lorentz-violation.| Substituting these relations back into the line element, the complete asymptotically (A)dS black hole metric for Case A takes the form
\begin{align}
    ds^2 &= -\left[1 - \frac{2M}{r} - \frac{(1 - \ell)\Lambda_e}{3} r^2\right] dt^2 \nonumber \\
    &+ \frac{(1 - \ell) dr^2}{1 - \frac{2M}{r} - \frac{(1 - \ell)\Lambda_e}{3} r^2} + r^2 d\theta^2 + r^2 \sin^2 \theta d\phi^2.
\end{align}
This solution reduces to the Schwarzschild-(A)dS solution when $ \ell = 0 $ and to its asymptotically flat counterpart when $ \Lambda_e = 0 $.

For Case B ($ C_1 = 1 $), the metric function becomes
\begin{equation}
A(r) = \frac{1}{1 - \ell} - \frac{2M}{r} - \frac{(b^2 \lambda + \Lambda)}{3(1 + \ell)} r^2,
\end{equation}
with $ \Lambda = \Lambda_e $. These solutions reduce to Schwarzschild-(A)dS when $ \ell = 0 $ and to Schwarzschild when $ \Lambda_e = 0 $, with $ \ell $, $ \Lambda_e $, and $ b^2 $ governing deviations from GR. Once again, invoking the relationship depicted in Eq. \eqref{e3}, the asymptotically (A)dS black hole metric for Case B becomes
\begin{align}
    ds^2 &= -\left(\frac{1}{1 - \ell} - \frac{2M}{r} - \frac{\Lambda_e}{3} r^2\right) dt^2 \nonumber \\
    &+ \frac{dr^2}{\frac{1}{1 - \ell} - \frac{2M}{r} - \frac{\Lambda_e}{3} r^2} + r^2 d\theta^2 + r^2 \sin^2 \theta d\phi^2.
\end{align}
This solution coincides with the previously known Schwarzschild-(A)dS-like black hole in KR gravity and, as in Case A, reduces to familiar results in the absence of Lorentz-violation or a cosmological constant.

A salient feature of these solutions lies in their distinct asymptotic structures. Although the leading asymptotic behaviors of the Ricci tensor components differ between Cases A and B, reflecting their disparate metric configurations, the Kretschmann scalar's corrections due to the cosmological constant are structurally similar in both cases. Specifically, the contribution from the cosmological constant modifies the Kretschmann scalar with a universal additive term proportional to $ \Lambda_e^2 $, modulated by the Lorentz-violating parameter $ \ell $.

Moreover, the horizon structure and existence of black holes depend sensitively on the sign and magnitude of $ \Lambda_e $. When $ \Lambda_e < 0 $ (the AdS case), black hole solutions exist for any parameter values, with a single event horizon enclosing the singularity. However, for $ \Lambda_e > 0 $ (the dS case), black hole solutions are restricted to specific regions of the $ (\ell, M^2\Lambda_e) $ parameter space. In Case A, the requirement for horizon formation is $ 9(1 - \ell)M^2\Lambda_e \geq 1 $, whereas in Case B, the stricter condition $ 9(1 - \ell)^3 M^2\Lambda_e \geq 1 $ must be satisfied. These inequalities demarcate regions in parameter space where the event and cosmological horizons exist and where the spacetime remains physically meaningful.

\section{Analytical considerations on the horizon formations} \label{sec3}
In this section, we derive the general analytic expression for the event horizon that is valid for the two cases mentioned in the previous section. We use a dummy variable $\zeta^2$ and Eq. (3) for brevity. That is, for Case A, $\zeta^2 = 1$, and $\Lambda = (1 - \ell) \Lambda_e$; for Case B, $\zeta^2 = 1/(1 - \ell)$, and $\Lambda = \Lambda_e$. These choices simplify the metric functions for each case while ensuring $\zeta > 0$, as $\ell < 1$ from the allowed parameter range (see Section IV). The reason for this choice of a dummy variable is to prevent $\zeta$ from becoming negative, which makes the black hole metric ill-defined \cite{Majumder:2024mle}.

The horizon radius of the dS case can be found by solving $r$ in the equation
\begin{equation}
    \left(\zeta^2 - \frac{2M}{r} - \frac{\Lambda r^2}{3}\right)/\chi^2 = 0.
\end{equation}
Rewriting the above equation into a neat form yields
\begin{equation} \label{e8}
    r^3 - \frac{3 \zeta^2}{\Lambda}\, r + \frac{6M}{\Lambda} = 0,
\end{equation}
which is a depressed cubic equation of the form $r^3 + pr + q = 0,$ with the identifications
\begin{equation}
    p = -\frac{3 \zeta^2}{\Lambda} \quad\text{and}\quad q = \frac{6M}{\Lambda}.
\end{equation}
In general, Cardano's method allows us to express the solution in trigonometric form for depressed cubics with three real roots (which happens when the discriminant is negative),
\begin{equation}
    r = 2\sqrt{-\frac{p}{3}}\, \cos\!\left(\frac{1}{3}\arccos\!\left(\frac{3q}{2p}\sqrt{-\frac{3}{p}}\right) - \frac{2\pi k}{3}\right),\quad k=0,1,2.
\end{equation}
The branches $k=0,1,2$ correspond to the cosmological horizon, event horizon, and a non-physical negative root. With slight back substitution and algebra, we find that
\begin{equation} \label{e11}
    r_{\rm h}^{\rm dS} = 2 \zeta\sqrt{\frac{1}{\Lambda}}\, \cos\!\left[\frac{1}{3}\arccos\!\left[\frac{3M\sqrt{\Lambda}}{\zeta^{3}}\right] + \frac{\pi k}{3}\right],
\end{equation}
where we have used the identity \(\arccos(-x) = \pi - \arccos(x)\), which is useful for physical interpretations. Note that the expression is valid provided the argument of the \(\arccos\) function, namely \( \frac{3M\sqrt{\Lambda}}{(1-\ell)^{3/2}} \), lies within the interval \([-1,1]\) so that all three roots are real.

\begin{figure}[htp!]
    \centering
    \includegraphics[width=\columnwidth]{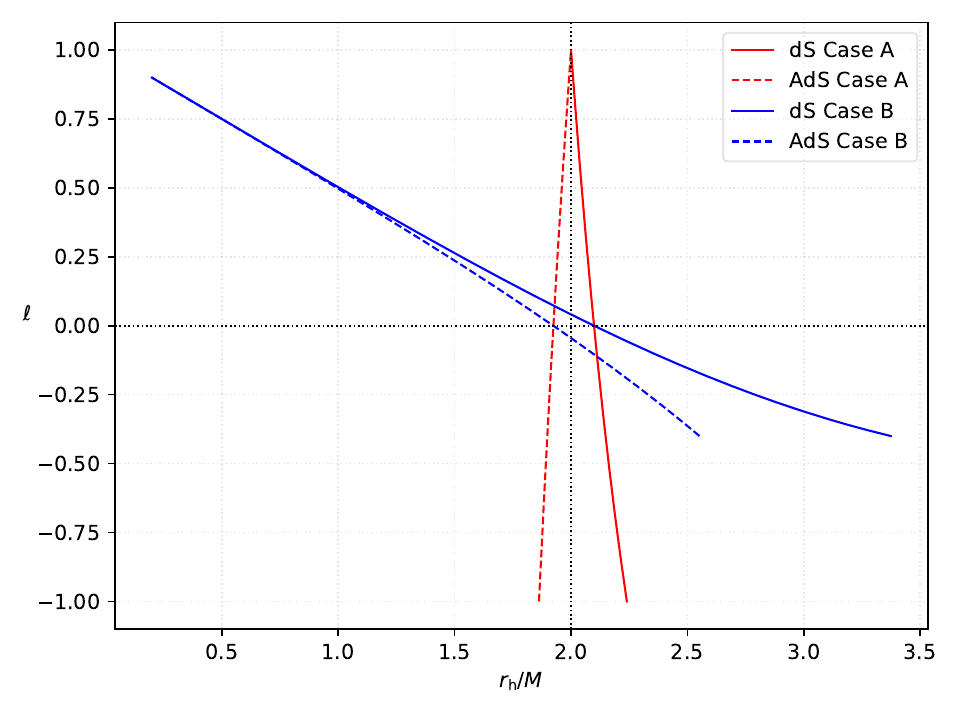}
    \caption{Behavior of the horizon for Case A and B. Here, we set $M = 1$ and scaled the cosmological constant as $\Lambda_e = 0.032$ to see the immediate effect. The vertical dotted line is added and corresponds to the Schwarzschild case for the sake of comparison.}
    \label{fig1}
\end{figure}

In Eq. \eqref{e11}, we cannot simply set $\Lambda$ to be negative if we want to study the horizon of the (A)dS case duly because the argument in the \(\arccos\) will become imaginary. By letting $\Lambda \rightarrow -|\Lambda|$, recast Eq. \eqref{e8} as
\begin{equation} 
    r^3 + \frac{3\zeta^2}{|\Lambda|}\, r - \frac{6M}{|\Lambda|} = 0.
\end{equation}
Unlike the de Sitter case, here \(p\) is positive, and the discriminant \(\Delta = -(4p^3 + 27q^2)\) is negative, so there is one real root and a pair of complex conjugate roots. In such cases, the unique real solution is best written in terms of hyperbolic functions. Using the standard formula for a depressed cubic when there is a single real solution, we write
\begin{equation}
    r = 2\sqrt{\frac{p}{3}}\,\sinh\!\Biggl[\frac{1}{3}\operatorname{asinh}\!\Bigl(-\frac{3q}{2p}\sqrt{\frac{3}{p}}\Bigr)\Biggr].
\end{equation}
With a little bit of back substitution, the unique real solution for the horizon radius is
\begin{equation} \label{e14}
    r_{\rm h}^{\rm (A)dS} = 2\zeta\sqrt{\frac{1}{|\Lambda|}}\;\sinh\!\Biggl[\frac{1}{3}\operatorname{asinh}\!\left[\frac{3M\sqrt{|\Lambda|}}{\zeta^{3}}\right]\Biggr].
\end{equation}
Note that from Eqs. \eqref{e11} and \eqref{e14}, we can immediately see that $\ell$ can only take values from the interval $(-\infty, 1]$. We plotted numerically the results as shown in Fig. \ref{fig1}.

\section{Photon sphere and Shadow analysis} \label{sec4}
The photon sphere $r_{\rm ps}$, is pivotal to shadow formation as it delineates the region where photons can orbit unstably, influencing the critical impact parameter that defines the shadow's edge. This boundary governs gravitational lensing: photons skimming outside the sphere are deflected to form the bright photon ring, while those crossing inside are captured, creating the dark silhouette observed. The sphere's role in setting the scale and sharpness of this optical feature arises from its control over light trajectories, with the ring's intensity enhanced by photons completing multiple orbits before escaping, making the photon sphere a fundamental determinant of the shadow's observable structure in GR \cite{Wei:2017mwc}.

Using a method that is widely known, the radius of the photon sphere can be determined by solving $r$ in the equation
\begin{equation} \label{e15}
	A(r)C'(r) - A'(r)C(r) = 0,
\end{equation}
where the prime denotes the differentiation with respect to $r$. Keeping in mind of the dummy variable $\zeta$ and Eq. \eqref{e3}, Eq. \eqref{e15} can be written as
\begin{equation} 
	3M - \zeta^2 r = 0,
\end{equation}
showing independence on $\Lambda$. Then, solving for $r$ gives
\begin{equation} 
	r_{\rm ps} = \frac{3M}{\zeta^2}.
\end{equation}
One can immediately verify that for Case A, the photon sphere is unaffected by both the parameters $\ell$ and $\Lambda_e$. Hence, $r_{\rm ps} = 3M$. For Case B, however,
\begin{equation} 
	r_{\rm ps}^{\rm B} = 3M(1-\ell).
\end{equation}

The critical impact parameter is defined as
\begin{equation}
	b_{\rm crit}^2 = \frac{C(r_{\rm ps})}{A(r_{\rm ps})},
\end{equation}
where $r_{\rm ps}$ is considered as the closest approach. Then, the general expression to accommodate Cases A and B is
\begin{equation}
	b_{\rm crit}^2 = \frac{27 M^{2}}{\zeta^{6}-9 \Lambda  \,M^{2}}.
\end{equation}

Referring to the method in Ref. \cite{Perlick:2021aok}, we can now obtain the exact expression for the invisible shadow radius as
\begin{equation} \label{e21}
	R_{\rm sh} = \left[ \left( \frac{27 M^{2}}{\zeta^{6}-9 \Lambda  \,M^{2}} \right) \left(\zeta^2 -\frac{2 M}{r_{o}}-\frac{\Lambda  \,r_{o}^{2}}{3}\right) \right]^{1/2}.
\end{equation}
We apply two approximations based on a realistic scenario, which is often useful in finding constraints using the EHT data. That is, we apply $\Lambda \rightarrow 0$ and $r_{o} \rightarrow 0$ on Eq. \eqref{e21} and find
\begin{equation} \label{e22}
	R_{\rm sh} \sim \frac{3 \sqrt{3}\, M}{\zeta^2} - \frac{\sqrt{3}\, r_o^{2} \Lambda  M}{2 \zeta^{4}} + \mathcal{O}(M^2).
\end{equation}
For Case A and Case B, we then have the following
\begin{equation}
	R_{\rm sh}^{\rm A} \sim 3 \sqrt{3}\, M - \frac{\sqrt{3}\, r_o^{2} (1-\ell)\Lambda_e  M}{2} + \mathcal{O}(M^2),
\end{equation}
\begin{equation}
	R_{\rm sh}^{\rm B} \sim 3 \sqrt{3}\, M (1-\ell) - \frac{\sqrt{3}\, r_o^{2} (1-\ell)^2 \Lambda_e  M}{2} + \mathcal{O}(M^2).
\end{equation}

Next, let us find some constraints on $\ell$ by using the EHT results. The black hole shadow's Schwarzschild deviation of Sgr A* ($2\sigma$ level) and M87* ($1\sigma$ level) are $4.209M \leq R_{\rm Schw} \leq 5.560M$ \cite{Vagnozzi:2022moj} and $ 4.313M \leq R_{\rm Schw} \leq 6.079M$ \cite{EventHorizonTelescope:2021dqv}, respectively. Let $\delta$ denotes the deviation from the standard Schwarzschild shadow radius, then for Sgr A*, $-0.364 \leq \delta/M \leq 0.987$. For M87*, $\delta/M = \pm 0.883$. These being said, we use Eq. \eqref{e22} and equate it to $3\sqrt{3} M + \delta$, which results to two solutions for constraint on the parameter $\ell$, applicable for Cases A and B:
\begin{equation} \label{e26}
    \zeta_1 \sim \pm \frac{3 \sqrt{M}}{\sqrt{\sqrt{3}\, \delta +9 M}} \mp \frac{\sqrt{\sqrt{3}\, \delta +9 M}\, r_o^{2} \Lambda}{36 \sqrt{M}},
\end{equation}
and
\begin{equation} \label{e27}
    \zeta_2 \sim \pm \frac{\sqrt{6\Lambda}\, r_o}{6}.
\end{equation}
In finding for the constraints in $\ell$ (which gives the upper and lower bounds), we find that Eq. \eqref{e15} gives the same outcome for Cases A and B, which do not depend on the deviation $\delta$:
\begin{equation}
    \ell_2^{\rm A,\,B} \sim 1 - \frac{6}{\Lambda_e r_o^2}.
\end{equation}
Using Eq. \eqref{e26}, however, we find differing constraints that consist of two pairs of upper and lower bounds for $\ell_1$:
\begin{align}
    \ell_1^{\rm A} &\sim 1 + \frac{\pm 36 \sqrt{\sqrt{3}\, \delta +9 M}\, \sqrt{M}-108 M}{\left(\sqrt{3}\, \delta +9 M \right) r_o^{2} \Lambda_e}, \nonumber \\
    \ell_1^{\rm B} &\sim \pm \frac{\sqrt{3}\, \delta}{9 M} - \frac{r_o^{2} \left(6 M \sqrt{3}\, \delta +27 M^{2}+\delta^{2}\right) \Lambda_e}{162 M^{2}}.
\end{align}

Table \ref{tab1} summarizes the numerical constraints for $\ell$ in both cases A and B. Interestingly, for Case A, the bounds on $ \ell $ spanned extremely large and asymmetric values suggesting sensitivity to highly nonlinear contributions and potential amplification of small deviations in the shadow radius formula. In contrast, Case B produced more symmetric and observationally plausible bounds, typically within $ |\ell| < 0.2 $, consistent with modest LSB. These results emphasize that Case B is more viable when matched against realistic astrophysical data and lends itself to phenomenologically relevant interpretations, while Case A, though mathematically consistent, may be observationally less favored without further constraint regularization.

While both Case A and Case B represent mathematically consistent black hole solutions in KR gravity, their contrasting constraints on the Lorentz-violating parameter  $\ell$  highlight differences in observational behavior rather than correctness. The large values of  $\ell$ inferred in Case A are not indicative of theoretical inconsistency but rather arise from the nonlinear dependence of the metric on $\ell$, which amplifies observational uncertainties, particularly from EHT measurements of the shadow radius. That is, the metric's high sensitivity to 
$\ell$ makes it hard to constrain because small observational noise leads to some large parameter uncertainty. In contrast, Case B introduces $\ell$ through a more linearly sensitive metric structure, resulting in tighter and more symmetric constraints that align well with Solar System tests and weak gravitational lensing. While the magnitude of $\ell$ alone is not a definitive criterion for model preference, Case B emerges as more favorable due to its stable parameter sensitivity, compatibility with multiple observables, and growing support in the literature. Therefore, under current observational precision, Case B offers a more practical and robust framework for probing Lorentz-violating effects in black hole spacetimes.

\begin{table}
\centering
\begin{tabular}{c|cccc}
\toprule
Solution & CASE & SMBH & $\delta$ & $\ell$ \\
\midrule
\multirow{8}{*}{$\zeta_1$} 
    & \multirow{4}{*}{A:} 
         & \multirow{2}{*}{Sgr A*} & $-0.364$ & $-6.380\times 10^{10}, -3.514\times 10^{12} $ \\[1ex]
    &                         &                        & $0.987$  & $1.270\times 10^{11}, -2.923\times 10^{12}$ \\[1ex]
    &                         & \multirow{2}{*}{M87*}   & $-0.883$ & $-4.326\times 10^{4}, -9.300\times 10^{5}$ \\[1ex]
    &                         &                        & $0.883$  & $2.818\times 10^{4}, -7.185\times 10^{5}$ \\
\cmidrule(lr){2-5}
    & \multirow{4}{*}{B:} 
         & \multirow{2}{*}{Sgr A*} & $-0.364$ & $-0.0705,0.0705$ \\[1ex]
    &                         &                        & $0.987$  & $0.1899,-0.1899$ \\[1ex]
    &                         & \multirow{2}{*}{M87*}   & $-0.883$ & $-0.1699,0.1699$ \\[1ex]
    &                         &                        & $0.883$  & $0.1699,-0.1699$ \\
\midrule
\multirow{2}{*}{$\zeta_2$} 
    & \multirow{2}{*}{A:} & SgrA  & ---      & $-8.318\times 10^{11}$ \\[1ex]
    &                         & M87   & ---      & $-2.019\times 10^{5}$ \\
\bottomrule
\end{tabular}
\caption{Summary of measurements for $\ell_1$ (Case A and Case B of metric) and $\ell_2$ (applicable for Case A only).}
\label{tab1}
\end{table}

\section{Deflection angle in weak field regime} \label{sec5}
To analyze the weak deflection angle, we apply the non-asymptotic form of the Gauss-Bonnet theorem, incorporating the finite-distance effects of both the source (S) and the receiver (R) \cite{Li:2020wvn}:
\begin{equation} \label{wda_Li}
    \Theta = \iint_{_{r_{\rm ps}}^{\rm R }\square _{r_{\rm ps}}^{\rm S}}KdS + \phi_{\text{RS}}.
\end{equation}

For a more generalized approach, the Jacobi metric is employed to account for the deflection angle of particles with mass $\mu$:

\begin{equation} \label{Jac_met}
    dl^2=g_{ij}dx^{i}dx^{j}
    =(E^2-\mu^2A(r))\left(\frac{B(r)}{A(r)}dr^2+\frac{C(r)}{A(r)}d\phi^2\right).
\end{equation}
Here, $E$ is the particle's total energy per unit rest mass, and $v$ represents the velocity of the particle as a fraction of the speed of light $c$. For massless particles like photons, where $v=1$, the energy is normalized as $E$. For massive particles $(v<1)$, the energy is expressed as $E = (1-v^2)^{-1/2}$. As shown in \cite{Li:2020wvn}, by selecting the photon sphere radius $r_{\rm ps}$ as one of the integration limits in Eq. \eqref{wda_Li}, the weak deflection angle for both massive and massless particles satisfies:
\begin{equation} \label{wda_Li2}
    \Theta = \int^{\phi_{\rm R}}_{\phi_{\rm S}} \int_{r_{\rm ps}}^{r(\phi)} K\sqrt{g} \, dr \, d\phi + \phi_{\rm RS}.
\end{equation}
In this equation, $\phi_{\rm RS} = \phi_{\rm R} - \phi_{\rm S}$, with $\phi_{\rm R} = \pi - \phi_{\rm S}$. The term $K$ denotes the Gaussian curvature \cite{Gibbons:2008rj,Ishihara:2016vdc}, while $g$ represents the determinant of the Jacobi metric, expressed as ($\Gamma_{rr}^{\phi} = 0$ from Eq. \eqref{Jac_met}):
\begin{equation} \label{G_curva}
    K=-\frac{1}{\sqrt{g}}\left[\frac{\partial}{\partial r}\left(\frac{\sqrt{g}}{g_{rr}}\Gamma_{r\phi}^{\phi}\right)\right],
\end{equation}
where 
\begin{equation}
    g = \frac{(E^2 - \mu^2 A(r))B(r)C(r)}{A(r)^2}.
\end{equation}
Consequently, when evaluating at $r_{\rm ps}$ \cite{Li:2020wvn},
\begin{equation}
    \left[\int K\sqrt{g}dr\right]\bigg|_{r=r_{\rm ps}} = 0,
\end{equation}
which leads to the integral relation:
\begin{align} \label{gct}
    &\int_{r_{\rm ps}}^{r(\phi)} K\sqrt{g}dr = \nonumber \\
    &-\frac{A(r)\left(E^{2}-A(r)\right)C'-E^{2}C(r)A(r)'}{2A(r)\left(E^{2}-A(r)\right)\sqrt{B(r)C(r)}}\bigg|_{r = r(\phi)}.
\end{align}

To evaluate the integral in Eq. \eqref{gct}, we must employ the orbit equation, which follows from the upper integration limit. Defining the inverse radial coordinate as $u = r^{-1}$, we obtain:
\begin{align}
    &F(u) \equiv \left(\frac{du}{d\phi}\right)^2 \nonumber \\
    &= \frac{C(u)^2u^4}{A(u)B(u)}\Bigg[\left(\frac{E}{J}\right)^2-A(u)\left(\frac{1}{J^2}+\frac{1}{C(u)}\right)\Bigg],
\end{align}
where $J = Evb$ represents the particle's angular momentum, and $b$ is the impact parameter. To simplify notation, we retain the variables $E$ and $J$. This leads to:
\begin{equation} \label{eorb}
    \left(\frac{du}{d\phi}\right)^2 = \frac{\frac{E^{2}}{J^{2}}-\left(\zeta^{2}-2 m u -\frac{\Lambda}{3 u^{2}}\right) \left(\frac{1}{J^{2}}+u^{2}\right)}{\chi^2}.
\end{equation}
The point of closest approach $u(\phi)$ can be found by imposing again the circular orbit condition, $\left(\frac{du}{d\phi}\right)^2 = 0$, yielding:
\begin{equation} \label{eorb2}
    u = \frac{1}{b}-\frac{\zeta -1}{v^{2} b}.
\end{equation}

We assume $\zeta$ is slightly above or below unity not only to preserve the Newtonian limit but also to avoid horrendously complicated approximation results \cite{Pantig:2024kqy}. We take the derivative of Eq. \eqref{eorb} with respect to $\phi$ and solve the differential equation, then with the help of Eq. \eqref{eorb2} and following the iterative method, we find the expression for the photon's closest approach as
\begin{align} \label{cl_app}
    &u(\phi) = \frac{1}{b}\sin\left( \frac{\zeta \phi}{\chi} \right) + \frac{m}{b^{2} v^{2}} \left[ 1+v^2\cos\left( \frac{\zeta \phi}{\chi} \right) \right] \nonumber \\
    &- \frac{\left(\zeta - 1\right) \left(b +4 m \right)}{b^{2} v^{2}} + \frac{b \Lambda}{6 v^{2}} \left[ 1 - \frac{\left(4 v^{2}-3\right) \left(\zeta - 1\right)}{v^{2}} \right]
\end{align}
After integration,
\begin{align} \label{e_int}
    \int_{\phi_{\rm S}}^{\phi_{\rm R}} &\int_{r_{\rm ps}}^{r(\phi)} K\sqrt{g} \, dr \, d\phi \sim -\frac{\left(2 E^{2}-1\right) m}{\left(E^{2}-1\right) b}\cos \! \left(\frac{\phi}{\chi}\right) \bigg\vert_{\phi_{\rm S}}^{\phi_{\rm R}} \nonumber \\ 
    &-\frac{\phi_{\rm RS}}{\chi}-\frac{\left(\zeta -1\right) \phi_{\rm RS}}{\chi} + \frac{b^{2} \left(E^{2}+1\right) \Lambda }{6 \,(E^{2}-1)}\cot \! \left(\frac{\phi}{\chi}\right) \bigg\vert_{\phi_{\rm S}}^{\phi_{\rm R}} \nonumber \\ 
    &+ C + \mathcal{O}\left[\left(\zeta-1\right)m, \left(\zeta-1\right)\Lambda_e, \left(\zeta-1\right)m \Lambda_e \right],
\end{align}
where $C$ is a temporary integration constant. From here on, we remove the higher order terms $\mathcal{O}$ for brevity. Finally, the weak deflection angle for massive particles is given by:
\begin{widetext}
\begin{align} \label{wda_gen}
    \Theta &\sim \frac{2m \left(v^{2}+1\right) }{v^{2} b}  \Bigg[ \sqrt{1-b^{2} u^{2}}+ \left( \arcsin \! \left(b u \right) - \frac{1}{v^{2} \sqrt{1-b^{2} u^{2}}} \right)bu(\zeta-1) \Bigg] \nonumber \\ 
    & -\Bigg\{\pi - 2 \Bigg[\arcsin \! \left(b u \right)+\frac{\zeta - 1}{v^{2} \sqrt{1-b^{2} u^{2}}}+\frac{\left[v^{2} \left(b^{2} u^{2}-1\right)-1\right] m}{\sqrt{1-b^{2} u^{2}}\, b \,v^{2}}-\frac{b^{2} \Lambda}{6 c \,v^{2} \sqrt{1-b^{2} u^{2}}} \Bigg] \Bigg\}\left( 1-\frac{\chi}{\zeta} \right) \nonumber \\
    & + \frac{b^{2} \left(v^{2}-2\right) \Lambda}{3 v^{2}} \left( \frac{\sqrt{1-b^{2} u^{2}}}{b u} \right).
\end{align}
\end{widetext}
In the far approximation where $u\rightarrow 0$, the above reduces to
\begin{align}
    \Theta^{\rm timelike} &\sim \frac{2 \left(v^{2}+1\right) m}{v^{2} b} - \Bigg[ \pi -\frac{2 \left(\zeta - 1\right)}{v^{2}}+\frac{2 \left(1+v^{2}\right) m}{b \,v^{2}} \nonumber \\
    &+\frac{b^{2} \Lambda}{3 \zeta \,v^{2}} \Bigg] \left( 1-\frac{\chi}{\zeta} \right) + \frac{b \left(v^{2}-2\right) \Lambda}{6 v^{2} u_o},
\end{align}
where $u_o = u_{\rm S} = u_{\rm R}$. Lastly, for photons $(v = 1)$,
\begin{align} \label{wda_null}
    \Theta^{\rm null} &\sim \frac{4m}{b} - \left( \pi -2 \zeta +2+\frac{4 m}{b}+\frac{b^{2} \Lambda}{3 \zeta} \right) \left( 1-\frac{\chi}{\zeta} \right) - \frac{b \Lambda}{6 u_o}.
\end{align}
We demonstrate Eq. \eqref{wda_null} on Cases A and B. For Case A,
\begin{align} \label{wda_nullA}
    \Theta^{\rm A} &\sim \pi(\sqrt{1-\ell}-1) + \frac{4m\sqrt{1-\ell}}{b} \nonumber \\
    &- b\Lambda_e (1-\ell) \left[ \frac{1}{6 r_o} - \frac{b(\sqrt{1-\ell}-1)}{3} \right].
\end{align}
For Case B,
\begin{align} \label{wda_nullB}
    &\Theta^{\rm B} \sim \pi(\sqrt{1-\ell}-1) + \frac{4m\sqrt{1-\ell}}{b} - \frac{b\Lambda_e}{6 r_o} \nonumber \\
    &- \frac{2\left[2\left(\sqrt{1-\ell}-1\right)+\ell\right]}{\sqrt{1-\ell}}-\frac{b\Lambda_e\left( \sqrt{1-\ell}+\ell-1 \right)}{3}
\end{align}

In the weak field regime, Solar System tests can effectively constrain parameters appearing in the weak deflection angle expression, as shown in Eq. \eqref{wda_null}. Within the parametrized post-Newtonian (PPN) formalism, the deflection angle of light is given by \cite{Chen:2023bao}:
\begin{equation} \label{ppn}
    \Theta^{\rm PPN} \backsimeq \frac{4M_{\odot}}{R_{\odot}}\left(\frac{n \pm \Delta}{2} \right),
\end{equation}
where $n = 1.9998$ and $\Delta = 0.0003$ \cite{Fomalont_2009}. The parameter $\Delta$ quantifies the observational uncertainty in measuring spacetime curvature due to the Sun's gravitational field, characterized by the solar mass ($M_\odot = 1476.61\,\text{m}$), observer distance ($r_o = 148.61\times10^{6}\,\text{m}$), and solar radius ($R_{\odot} = 6.96\times 10^{8}\,\text{m}$). The deflection angle $\Theta^{\rm PPN}$, expressed in radians, is closely associated with solar system measurements of light bending, particularly through astrometric data from instruments such as the Very Long Baseline Array (VLBA) \cite{Fomalont_2009}. By equating Eqs. \eqref{ppn} and \eqref{wda_null}, we derive a general relation for constraining the parameter $\chi$:
\begin{align}
    \chi &= \zeta + \frac{\chi R_\odot \Lambda}{6 r_o \left(-2 \chi +\pi +2\right)} + \frac{2 \chi M_\odot \left(n +\Delta -2\right)}{R_\odot \left(-2 \chi +\pi +2\right)} \nonumber \\
    &- \frac{2 \Lambda  \left(R_\odot \left(n +\Delta -2\right) r_o +\chi \right) M_\odot}{3 \left(2 \chi -\pi -2\right)^{2} r_o}.
\end{align}
Now, if we do consider Case A and taking note the substitution for what $\zeta$, and $\Lambda$ are, we obtain
\begin{align} \label{e48}
    \ell^{\rm A} &\sim -\frac{\Lambda_e  R_\odot}{3 \pi  r_o}-\frac{4 M_\odot \left(n-2 +\Delta \right)}{R_\odot \pi} \nonumber \\
    &+\frac{4 \left(3-n -\Delta +R_\odot r_o \left(n-2 +\Delta \right)\right) M_\odot \Lambda_e}{3 \pi^{2} r_o}.
\end{align}
We find the constraint of $\ell$ within the bounds $[-2.701\times10^{-10},1.351\times10^{-9}]$, demonstrating that even minute symmetry-breaking effects leave observable imprints on photon trajectories. These limits are remarkably consistent with prior post-Newtonian constraints on Lorentz-violating theories and affirm that even subtle symmetry-breaking effects imprint measurable signatures on photon trajectories at solar scales. Notably, while EHT-based constraints are broader due to their astrophysical context, Solar System observations provide far tighter limits on $ \ell $, reinforcing their complementary role in testing modified gravity. As a final remark, we found out that for Case B, the expression is the same as that of Eq. \eqref{e48}, except for the third term. We note that the dominant factor is the 2nd term for both cases, implying that the Solar System test cannot identify or distinguish the difference between these cases.

\section{Topological Thermodynamics and Photon Spheres}  \label{sec6}

\subsection{Topological Thermodynamics} \label{ssec6.1}
In recent years, investigations into critical phenomena and phase transitions of (A)dS black holes have greatly intensified, particularly within the framework of extended black hole thermodynamics. In this extended phase space, the negative cosmological constant \(\Lambda\) acquires a thermodynamic interpretation as the thermodynamic pressure \(P\), defined via the relation:
\begin{equation}
P = -\frac{\Lambda}{8 \pi G}.
\end{equation}
This perspective enriches our understanding of black hole thermodynamics, enabling analogies with traditional thermodynamic systems and phase transitions \cite{Dolan:2011xt,Kubiznak:2012wp,Kubiznak:2016qmn,Karch:2015rpa,Mancilla:2024spp}.

Recently, a novel viewpoint—thermodynamic topology—has been developed by Wei et al. based on Duan's topological current theory \cite{Wei:2022dzw}. This method characterizes critical points in the thermodynamics of black holes by associating them with topological charges arising from a topological current \(j^\mu\). Operationally, one constructs a scalar thermodynamic potential \(\Phi\), generally related to temperature, and analyzes a derived two-dimensional vector field whose zero points correspond precisely to thermodynamic criticalities. Each critical point carries a specific topological charge (\(+1\) or \(-1\)), and the total topological charge is the algebraic sum of individual contributions.

Complementarily, another topological approach characterizes black hole solutions as topological defects embedded within thermodynamic parameter spaces. By analyzing these defects and computing winding numbers, one can infer the stability properties of black hole solutions in a robust topological manner \cite{Gogoi:2023qku,Shahzad:2023cis,Shahzad:2024ojx,Shahzad:2024ycq,Shahzad:2024pti,Liu:2024axg,Zhu:2024jhw}.

To derive explicit expressions within this thermodynamic context, we first determine the mass \( M \) of our black hole solution by setting the metric function \( f(r) = 0 \) at the horizon radius \( r_+ \):
\begin{equation}
M = \frac{r_+}{2}\left(\frac{1}{1 - \ell} - \frac{\Lambda_e}{3} r_+^2\right),
\end{equation}
where \( r_h = r_+ \) denotes the event horizon radius. Expressing the effective cosmological constant \(\Lambda_e\) explicitly in terms of the thermodynamic pressure \(P\):
\begin{equation}
P = -\frac{\Lambda_e}{8\pi}, \quad \Rightarrow \quad \Lambda_e = -8\pi P,
\end{equation}
we arrive at the mass explicitly in terms of thermodynamic parameters as:
\begin{equation}
M = \frac{r_+}{2}\left(\frac{1}{1 - \ell} + \frac{8\pi P r_+^2}{3}\right).
\end{equation}
This expression identifies how the mass \(M\) depends explicitly on horizon radius \(r_+\), parameter \(\ell\), and pressure \(P\), emphasizing its direct thermodynamic significance.

Subsequently, the black hole temperature \(T\) can be computed from standard thermodynamic identities as:
\begin{equation}
T=\frac{\partial_{r_{+}} M}{\partial_{r_{+}} S} = \frac{8\pi P r_+^2 (1 - \ell) - 1}{4\pi r_+ (1 - \ell)}.
\end{equation}
To identify the critical points, we set the second derivative of temperature with respect to entropy (or equivalently radius) to zero:
\begin{equation}
\frac{\partial_{r_{+}} T}{\partial_{r_{+}} S}=0,\quad\Rightarrow\quad P=\frac{1}{8\pi\,(\ell-1)\,r_+^2}.
\end{equation}
Using this relation, we obtain a simplified expression for the critical temperature:
\begin{equation}
T = \frac{1}{2\pi\,r_+\,(\ell-1)}.
\end{equation}
This compact form highlights the explicit dependence of critical temperature on the horizon radius \(r_+\) and parameter \(\ell\).

Following Duan's topological procedure, we now define the scalar potential \(\Phi\) as \cite{Cunha:2020azh,Wei:2020rbh,Wei:2021vdx,Hosseinifar:2024wwe,Chen:2024atr}:
\begin{equation}
\Phi(r_+, \theta) = \frac{T}{\sin \theta}.
\end{equation}
This scalar potential allows us to systematically analyze thermodynamic criticalities by constructing a two-dimensional vector field \(\phi=(\phi_{r_+},\phi_{\theta})\) from the gradients of \(\Phi\). The radial and angular components of the vector field are explicitly computed as:
\begin{equation}
\phi_{r_+} 
= \frac{\partial \Phi}{\partial r_+} 
= \frac{\csc (\theta )}{2 \pi  r_+^2(1 - \ell)},\quad
\phi_{\theta} 
= \frac{\partial \Phi}{\partial \theta} 
= \frac{\cot (\theta ) \csc (\theta )}{2 \pi  r_+(1 - \ell)}.
\end{equation}

To perform a detailed topological analysis, we must normalize the vector field:
\begin{equation}
n_{r_+} = \frac{\phi_{r_+}}{\sqrt{\phi_{r_+}^2 + \phi_{\theta}^2}}, \quad
n_{\theta} = \frac{\phi_{\theta}}{\sqrt{\phi_{r_+}^2 + \phi_{\theta}^2}}.
\end{equation}
Explicitly, these normalized components become:
\begin{widetext}
\begin{equation}
n_{r_+} = \frac{\csc(\theta)}{\sqrt{\csc^2(\theta)+\cot^2(\theta)}},\quad
n_{\theta} = \frac{\cot(\theta)\csc(\theta)}{\sqrt{\csc^2(\theta)+\cot^2(\theta)}}.
\end{equation} 
\end{widetext}

\begin{figure}[htp!]
    \centering
\includegraphics[width=\columnwidth]{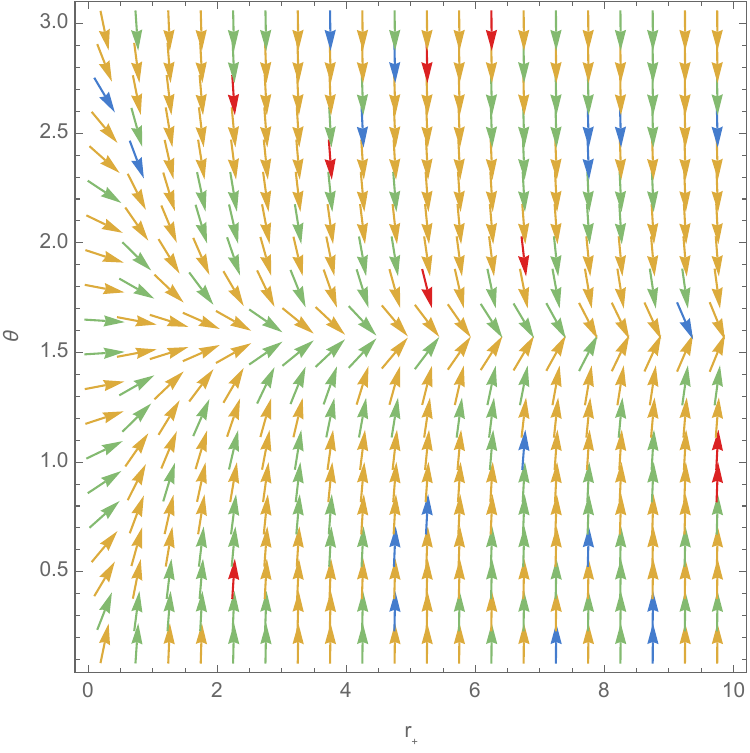}
    \caption{Normalized vector field \(n\) in the plane of horizon radius \(r_+\) versus \(\theta\) for an (A)dS black hole with parameter \(\ell=0.2\). Field line behavior around critical points provides direct visual insight into the system's topological thermodynamics.}
    \label{fig:vector1}
\end{figure}

Figure \ref{fig:vector1} illustrates the normalized vector field \(n\) in the \((r_+,\theta)\) plane for the (A)dS black hole in the canonical ensemble, setting the parameter \(\ell = 0.2\). This visualization reveals the geometric structure and stability properties of the black hole thermodynamics. The divergence (convergence) of field lines around critical points indicates a topological charge of \(+1\) (\(-1\)), respectively. Our analysis, however, demonstrates that no critical points occur precisely at \(\theta = \pi/2\), indicating that topological thermodynamics, in this particular setup, does not yield any additional critical information at this angular position.

Finally, to quantify the topological charges explicitly, we compute the deflection angle \(\Omega\) of the vector field along a closed contour \(C\):
\begin{equation}
\Omega(\vartheta)=\int_0^\vartheta \epsilon_{ab}n^a\partial_\vartheta n^b  d\vartheta,\quad Q=\frac{1}{2\pi}\Omega(2\pi),
\end{equation}
by introducing appropriate parametrization:
\begin{align}
\nonumber
r_+ &= a \cos\vartheta +r_{0},\\
\theta &= b \sin\vartheta+\frac{\pi}{2}.
\end{align}
This approach circumvents singularities at \(\theta = \pi/2\), allowing a robust numerical evaluation of the topological charges.
This topological analysis clarifies the intrinsic relationship between black hole thermodynamics and photon sphere topology, offering a powerful geometric perspective on critical phenomena and stability properties in gravitational thermodynamics.

\subsection{Topological Photon Sphere} \label{ssec6.2}
Next, we explore the topological properties of the photon sphere surrounding our black hole solution. To initiate this discussion, we introduce a regular potential function, inspired by previous works \cite{Cunha:2017qtt,Cunha:2020azh,Wei:2020rbh, Sadeghi:2024itx}, defined as follows:
\begin{equation} \label{15}
H(r, \theta)=\sqrt{\frac{-g_{tt}}{g_{\varphi\varphi}}}=\frac{1}{\sin\theta}\left(\frac{f(r)}{h(r)}\right)^{1/2}.
\end{equation}

The physical interpretation of this potential $H(r,\theta)$ is essential, as it provides insights into the geometry and stability of photon orbits around the black hole. Specifically, the photon sphere corresponds to the critical radius $r_{ps}$, satisfying the condition \begin{equation} \partial_{r}H=0. \end{equation} This indicates the photon orbits are stationary points of the potential, enabling photons to move along circular paths at this precise radius.

To further probe the topological structure around the photon sphere, we define a two-component vector field $\phi=(\phi^r,\phi^\theta)$ as: \begin{equation}\label{16}
\begin{split}
&\phi^r=\frac{\partial_rH}{\sqrt{g_{rr}}}=\sqrt{g(r)}\partial_{r}H,\\
&\phi^\theta=\frac{\partial_\theta H}{\sqrt{g_{\theta\theta}}}=\frac{\partial_\theta H}{\sqrt{h(r)}}.
\end{split}
\end{equation}

This vector field encapsulates the topological features of the spacetime near the photon sphere. Importantly, the points where $\phi$ vanishes (critical points) coincide precisely with the location of photon spheres. Utilizing Duan's topological approach, one defines a topological current whose charge $Q$ arises solely from the zeros of the vector field. Thus, each zero point of $\phi$ can be assigned a topological charge $Q$ that corresponds directly to its winding number.

To calculate this winding number, consider a closed, smooth, positively oriented curve $C_i$ enclosing only the $i_{th}$ critical point. Then, the winding number is given by \begin{equation}\label{17} \omega_i=\frac{1}{2\pi}\oint_{C_{i}}d\Lambda,\quad\text{with}\quad \Lambda=\frac{\phi^2}{\phi^1}. \end{equation}

The total topological charge, combining all contributions from multiple critical points, is simply the sum of individual winding numbers: \begin{equation}\label{19} Q=\sum_{i}\omega_{i}. \end{equation}

Physically, a nonzero winding number indicates the presence of stable or unstable photon spheres, depending on whether the corresponding critical point of $H(r,\theta)$ is a minimum or maximum, respectively. Conversely, if a closed curve does not enclose any zero points, the resultant topological charge necessarily vanishes, indicating the absence of photon spheres within that domain.

Let us now specify the potential explicitly for our scenario, as studied previously \cite{Cunha:2020azh,Wei:2020rbh,Hosseinifar:2024wwe}: \begin{equation} H(r,\theta)=\frac{1}{\sin\theta}\sqrt{\frac{f(r)}{h(r)}}=\csc (\theta )\sqrt{-\frac{\frac{r}{\ell-1}+2 M}{r^3}-\frac{\Lambda_e}{3}}, \end{equation} with the choice $h(r)=r^2$. This explicit form allows for a precise analytical and numerical investigation of the photon sphere topology.

To analyze topological phase transitions associated with the photon sphere, we again define the vector components explicitly as: \begin{eqnarray} \phi_{r}=\sqrt{A(r)},\partial_r H(r,\theta),\quad \phi_{\theta}=\frac{1}{\sqrt{h(r)}}\partial_{\theta} H(r,\theta), \end{eqnarray} which, after algebraic simplifications, yield \begin{eqnarray} \phi_{r}=\frac{\csc (\theta )(3(\ell-1) M+r)\sqrt{-\frac{3}{\ell-1}-\frac{6M}{r}-\Lambda_e r^2}}{(\ell-1) r^4 \sqrt{-\frac{3}{(\ell-1) r^2}-\Lambda_e-\frac{6 M}{r^3}}} \\ \phi_{\theta}=-\frac{\cot(\theta)\csc(\theta)\sqrt{-\frac{\frac{r}{\ell-1}+2 M}{r^3}-\frac{\Lambda_e}{3}}}{r}. \end{eqnarray}

To ensure clarity in the topological investigation, we normalize these vectors by defining the unit vector field: \begin{equation} n_r=\frac{\phi_{r}}{||\phi||},\quad n_{\theta}=\frac{\phi_{\theta}}{||\phi||}\label{unitv}. \end{equation}

This normalized vector field $(n_r,n_\theta)$ allows an intuitive visualization of vector directions and their variations, clearly illustrating any topological phase transitions occurring around the photon sphere.

\begin{figure}[h] \centering \includegraphics[width=\columnwidth]{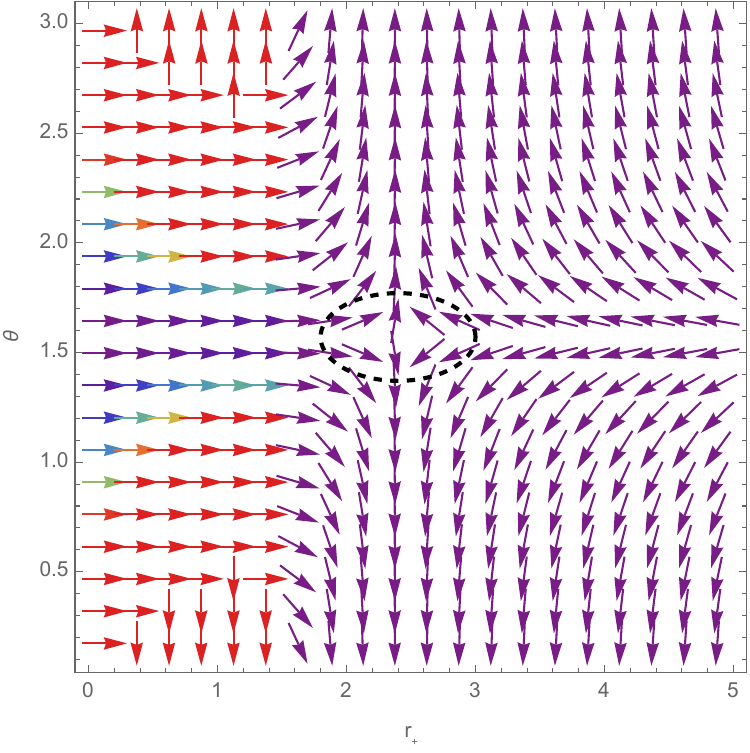} \caption{Normalized vector field $n$ plotted in the $(r,\theta)$ plane for an (A)dS-type black hole solution. The photon sphere emerges clearly at $(r,\theta)=(2.4, 1.57)$, computed for parameters $(\ell = 0.2, M = 1, \Lambda_e = 0.002)$. The zero-point singularity in the vector field indicates the presence and topological nature of the photon sphere.} \label{fig:vector2} \end{figure}

\begin{figure}[h] \centering \includegraphics[width=\columnwidth]{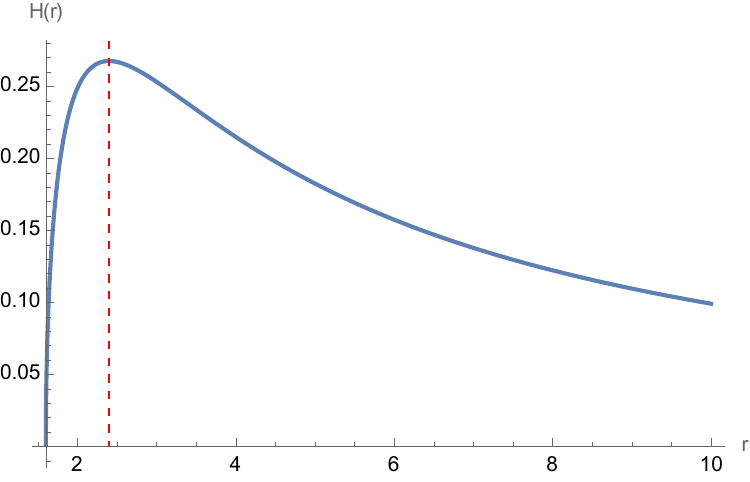} \caption{Topological potential $H(r)$ plotted versus radial coordinate $r$. This potential exhibits a clear global maximum precisely at the photon sphere radius $r_{ps}=2.4$, computed for the same parameters $(\ell = 0.2, M = 1, \Lambda_e = 0.002)$. Photons orbiting at this radius occupy an inherently unstable equilibrium position, from which minor perturbations will cause photons to either spiral inward or escape outward, reflecting observationally relevant unstable photon rings.} \label{fig:vector3} \end{figure}

Inspecting Fig. \ref{fig:vector2}, we observe the existence of a single zero point situated outside the event horizon. Consistent with prior results from the literature \cite{Wei:2020rbh}, this implies a topological total charge (TTC) of $-1$, signifying the inherent instability of the corresponding photon sphere.

In Fig. \ref{fig:vector3}, the potential $H(r)$ is depicted explicitly. The potential's pronounced global maximum at the photon sphere radius highlights its unstable character. Physically, any photon orbiting precisely at this radius experiences an unstable equilibrium: infinitesimal perturbations rapidly destabilize its orbit, resulting in photons spiraling towards or away from the black hole, a critical phenomenon directly linked to observational signatures such as photon rings and black hole shadows.

Thus, our topological analysis not only confirms the location of the photon sphere but also clarifies its physical nature as an inherently unstable orbit, a feature of significant observational and theoretical interest in contemporary gravitational physics.

\section{Conclusion} \label{conc}
In this work, we have conducted a comprehensive theoretical investigation of asymptotically (A)dS black holes within the framework of KR gravity, a Lorentz-violating theory motivated by string-theoretic considerations. Two distinct static and spherically symmetric black hole solutions were analyzed, referred to as Case A and Case B, each characterized by different couplings of the antisymmetric KR tensor field to the gravitational background. We derived general analytic expressions for the event horizon, photon sphere radius, shadow radius, and weak deflection angle for both dS and AdS spacetimes. These results exhibit a high degree of generality and are not only applicable within KR gravity but can also be extended to broader classes of non-rotating spacetimes. Importantly, we validated these expressions against current observational data from the EHT, particularly for the black holes Sgr A* and M87*, and through classical solar system lensing experiments using the Parameterized Post-Newtonian (PPN) framework. The comparison revealed that Case B offers more consistent and observationally viable bounds on the Lorentz-violating parameter $\ell$, with symmetric constraints tightly clustered around zero, whereas Case A displays sensitivity to nonlinear contributions that amplify deviations.

Beyond metric observables, we introduced topological diagnostics to analyze the thermodynamic and optical structure of the black hole solutions. In the thermodynamic context, we employed a topological current method to characterize phase transitions and stability in terms of winding numbers and topological charges derived from temperature-based vector fields. This revealed that the black hole mass, temperature, and pressure are intrinsically linked to the Lorentz-violating parameter $\ell$, leading to insights into critical behavior. In the optical regime, we analyzed the photon sphere topology using potential-based vector fields and demonstrated that the topological charge of the photon sphere is $-1$, confirming its unstable nature, which is a feature closely tied to the observed black hole shadow. Collectively, these multimodal approached, combining shadow analysis, deflection angle measurements, and topological field theory, provide a robust and unified platform for testing deviations from GR. Future work may extend this framework to include rotating or charged KR black holes, explore quasinormal modes and accretion dynamics, and assess detectability through gravitational wave observations. The results presented here underscore the viability of KR gravity as a testable extension of Einstein's theory and demonstrate the effectiveness of topological methods in decoding the deep structure of modified spacetimes.

In comparing our results with other black hole modifications involving Lorentz-violating fields, we note both synergies and distinctions. Maluf and Neves \cite{Maluf:2020kgf} derive static black hole solutions in bumblebee gravity with a nonzero cosmological constant, finding that Lorentz-violation modifies the horizon structure similarly to our Case A and Case B solutions. However, their vector field-induced Lorentz-violation yields tighter constraints on the symmetry-breaking parameter (e.g., $|\ell| \lesssim 10^{-14}$ from Solar System tests) compared to our KR field's antisymmetric tensor, where Case B constrains $|\ell| < 0.2$ from EHT data. Belchior et al. \cite{Belchior:2025xam} extend bumblebee gravity with a global monopole, introducing topological charges that resonate with our thermodynamic topology analysis in Section VI.A, though our KR framework avoids additional field couplings, simplifying the phase transition structure. In Ref. \cite{Fathi:2025byw}, the authors explored KR gravity with a self-interacting KR field and global monopole, reporting shadow radii deviations consistent with our Case B ($\sim 3\sqrt{3}M(1-\ell)$), but their model predicts stronger lensing effects due to the monopole's topological influence, unlike our cosmological constant-driven corrections. Baruah et al. \cite{Baruah:2025ifh} analyze quasinormal modes and Hawking radiation in KR gravity with a global monopole, finding thermodynamic stability properties that align with our Case B's topological charge analysis, though our focus on photon sphere stability (Sect. \ref{ssec6.2}) provides complementary optical insights. These comparisons tell that our Case B solution offers a robust, observationally viable framework for probing Lorentz-violation, with its simpler field structure facilitating tighter EHT-based constraints compared to models with additional topological or self-interacting fields, while maintaining consistency with Solar System tests.

\acknowledgments
R. P. and A. O. would like to acknowledge networking support of the COST Action CA18108 - Quantum gravity phenomenology in the multi-messenger approach (QG-MM), COST Action CA21106 - COSMIC WISPers in the Dark Universe: Theory, astrophysics and experiments (CosmicWISPers), the COST Action CA22113 - Fundamental challenges in theoretical physics (THEORY-CHALLENGES), the COST Action CA21136 - Addressing observational tensions in cosmology with systematics and fundamental physics (CosmoVerse), the COST Action CA23130 - Bridging high and low energies in search of quantum gravity (BridgeQG), and the COST Action CA23115 - Gravitational Quantum Physics and Metrology (Relativistic Quantum Information). A. O. also thank TUBITAK and SCOAP3, Turkiye for their support.

\bibliography{ref}

\end{document}